\documentclass[preprint,12pt]{elsarticle}

\usepackage{graphicx}
\usepackage{amssymb}

\graphicspath{{figures/}}
\journal{Physics of the Dark Universe}

\begin{document}

\begin{frontmatter}


\title{On the Direct Detection of Dark Matter in the Stratosphere }



%

\author[trieste,triesteu]{    G.~Cantatore*  }  
\author[freiburg,x1]{    H.~Fischer  }
\author[cern,x2]{    W.~Funk  }
\author[rijeka,cerijeka,trieste,x3]{   M.~Karuza } 
\author[messiah,x4]{   A.~Kryemadhi }
\author[patras,x5]{    M.~Maroudas}  
\author[capp,kaist,x6]{    Y.~K.~Semertzidis  }
\author[patras,x7]{    K.~Zioutas  } 
%
\address[capp]{Center for Axion and Precision Physics Research, Institute for Basic Science (IBS), Daejeon 34141, Republic of Korea.}
\address[kaist]{Department of Physics, Korea Advanced Institute of Science and Technology (KAIST), Daejeon 34141, Republic of Korea}
\address[freiburg]{Physikalisches Institut, Albert-Ludwigs-Universit\"{a}t Freiburg, 79104 Freiburg, Germany}
\address[cern]{European Organization for Nuclear Research (CERN), 1211 Gen\`eve, Switzerland}
\address[patras]{Physics Department, University of Patras, GR 26504,Patras, Greece}
\address[messiah]{Dept. Computing,Math \& Physics, Messiah University, Mechanicsburg PA 17055, USA}
\address[rijeka]{University of Rijeka, Department of Physics, Rijeka, Croatia.}
\address[cerijeka]{University of Rijeka, Photonics and Quantum Optics Unit, Center of Excellence for Advanced Materials and Sensing Devices, and Centre for Micro and Nano Sciences and Technologies, Rijeka, Croatia }
\address[trieste]{Istituto Nazionale di Fisica Nucleare (INFN), Sezione di Trieste, Trieste, Italy} 
\address[triesteu]{Universit\`a di Trieste, Trieste, Italy}
%
%

%
%
%
%
\cortext[mycorrespondingauthor]{Contact: Giovanni.Cantatore@cern.ch}

\fntext[x1]{Horst.Fischer@cern.ch}
\fntext[x2]{Wolfgang.Funk@cern.ch}
\fntext[x3]{mkaruza@phy.uniri.hr}
\fntext[x4]{akryemadhi@messiah.edu}
\fntext[x5]{marios.maroudas@cern.ch}
\fntext[x6]{yannis@bnl.gov}
\fntext[x7]{Konstantin.Zioutas@cern.ch}

%



\begin{abstract}
We investigate the prospects for direct detection of Dark Matter (DM) particles, such as dark photons, incident on the upper atmosphere. A recent work relates the burst-like temperature excursions in the stratosphere at heights of $\approx$38-47 km with incident invisible streaming matter. Surprisingly, dark photons match the reasoning presented in that work provided they constitute part of the local DM density. Dark photons mix with real photons with the same total energy without the need for an external field as would be required for instance for axions. Furthermore, the ionospheric plasma column above the stratosphere can enhance the dark photon to photon conversion due to resonance. 
The stratosphere is easily accessible via balloon flights.  
Balloon missions with up to a few tons of payload can be readily assembled to operate for 1-2 months at such atmospheric heights making for realistic short term endeavor of on-site direct DM search following this proposal. The approach need not be limited to streaming dark photons as other DM candidates might be searched simultaneously. 
Combining different types of measurements in a multi-band detector system and/or relating such investigations with other concurrent atmospheric observations, e.g. from space, may be proven to be the missing new approach in direct DM detection.
\end{abstract}


\end{frontmatter}


\section{Introduction} \label{S:1}
Since the first astrophysical observation of the “dunkle Materie” by Zwicky in 1933, we know that Dark Matter (DM) makes ~27\% of the matter-energy budget in the Universe. Its elemental composition however remains one of the greatest mysteries in all of physics. We recall that DM plays a major role in structure formation in the Universe. There is no lack of DM candidates, and therefore a large number of experimental searches are being undertaken to answer this physics question of fundamental importance. With the direct DM searches being in the forefront, also accelerator experiments (e.g. LHC experiments and other) and also other branches have targeted this question aiming to produce or detect some DM particle. Notably, the ultimate identification of DM species can be only with those who fill the Universe and are also expected in our neighbourhood.
In a recent investigation of the stratospheric temperature  excursions occurring annually around December-January, it was suggested that their planetary dependence points at streams of invisible matter from the dark universe as being the cause \cite{Zioutas:2020PDU}. In this work we go a step further and suggest how to detect them \textit{in situ}, i.e., at the stratosphere where the hours- or day-long heating-ups occur. As it was explained in \cite{Sirks:2020arxiv,Barak:2020arxiv,Zioutas:2020arxiv,Zioutas:2020PDU,Essig:2019arxiv,Meier:1997AP,Henry:2015ApJ}, the widely mentioned axions or WIMPs come not in question, at least not those which various experimental groups are searching for since some decades, since their interaction cross section with normal matter is far too small to account for the observed heating. Therefore, inspired by the reasoning about the origin of the stratospheric temperature anomalies, the stratosphere might be the place with occasionally enhanced converted DM constituents. We outline here a direct experimental search for DM candidates in the stratosphere, i.e., some 40 km above Earth’s surface. This novel idea of direct detection brings the experiment from underground or ground to the upper atmosphere.

In this proposal we focus primarily on dark photons, which are also DM candidates. The reasoning behind this choice is the following: dark photons can convert kinetically to ordinary photons, by contrast to the case with axions or axion-like particles, where a magnetic or electric field is needed to assist their photon conversion [3]. Notably, if the charge density has a plasma frequency which matches the rest mass of the dark photon, the conversion is enhanced due to resonance/coherence effects [3]. Following previous astrophysical observations combined with the axion or axion-like exclusion by the CAST experiment \cite{CAST:2017nat}, the dark photon rest mass might be up to the neV range. The corresponding plasma density is of the order of $10^{4 \pm 2}$ electrons/cm$^{3}$. Interestingly, such electron densities exist above the stratosphere (altitude above 100 km) [2]. Then, the dark photon scenario could fit-in for the heating up. Low speed streams of dark photons incident on the solar system experience solar and planetary gravitational self-focusing towards the Earth, before entering the ionosphere with the appropriate plasma density  some 100s to 1000s km above the Earth’s surface, where they can convert to real photons. The photons from resonant conversion would be RF photons in the MHz range. However, non-resonant conversions at other bands could be happening, albeit not the same enhancement as resonant ones. One such a region of interest is UV range around 6-8 eV because of the absorption edge in the upper stratosphere, which strongly prevents these photons from reaching ground level.

Thus, once an instrument, with its field-of-view pointing away from the Earth, observes temporally enhanced radio signals or UV photons in the stratosphere this will be the new direct DM signature, e.g., for converted dark photons or other ways indirectly. If a detected signature correlates with the local stratospheric temperature, and also exhibits matching planetary dependence, this strongly points to an invisible matter origin of the signal and at the same time excluding background sources.

Once such measurements in the stratosphere have been performed, they also could be correlated with missions in space including the International Space Station.

Such measurements could become the harbinger of a potential major direct discovery of DM constituents. Therefore, a multi-purpose detector system has the potential to unravel more species from the dark sector, which might well show enhanced fluxes in the upper atmosphere.

In short, we propose novel searches for DM constituents, in a location complementary to current DM searches and exploiting new signatures. 

The widely assumed annual oscillation of the DM wind is here overruled by planetary dependence \cite{Bertolucci:2017pdu}, which has not been so far considered in direct DM searches.

\section{The concept}
\label{S:2}
We suggest to directly search for streaming DM incident on the Earth’s atmosphere. Our motivation stems also from the fact that a possible signature of DM might be shielded by the atmosphere, in which case ground and underground DM search experiments would be quasi-blind. We propose as the detector location the stratosphere due to the planetary dependence of the temperature already observed there \cite{Zioutas:2020PDU,Essig:2019arxiv} and to the natural occurring plasma in the ionosphere above, which could produce coherent conversion effects.

Furthermore, it is known that solar photon irradiation of the atmosphere shows various characteristic absorption layers. The stratosphere around 40-45\,km above the surface absorbs UV~photons in the energy range $\sim6-8$ \,eV because of the overall oxygen absorption bands \cite{Meier:1997AP}. Thus, any search in this energy band should be done at stratospheric altitudes. 

We assume that incoming slow DM constituents first experience gravitational effects by the solar system bodies, including the Sun, and then are transformed into photons in the energy range mentioned above. Only photons emerging from slow DM constituents would show a planetary relationship. On the other hand, light or relativistic cosmic radiation cannot show a planetary dependence, since even the strongest possible gravitational focusing effect by the Sun has a focal length of $\sim 550$ Astronomical Units (A.U.) for photons.

A favored candidate would be the dark photon since it kinetically mixes with real photons without the need for electric or magnetic fields, and, in addition, this conversion could be resonantly enhanced if the plasma density matches the rest mass of the dark photon $(A^{'})$, $\hbar \omega_{plasma} = m_{A'}$ \cite{Gelmini:2020arxiv}.
The interaction of a dark photon with a plasma medium is described for example in \cite{Gelmini:2020arxiv,Dubovsky:2018arxiv}. 
The ionospheric plasma density is changing steadily, reaching a peak in electron density at an altitude of about 300 km. The typical electron density of the ionosphere is $N_e=10^{12}$\,electrons/m$^3$ resulting in a plasma frequency of

\begin{equation}
\label{eq:PlasmaFreq}
\omega_{plasma}=2\pi \sqrt{\frac{N_e e^2}{m_e \epsilon_0}}\approx 50 \mbox{[Hz]}\cdot \sqrt{N_e/ \mbox{m}^{-3}}=50\,\mbox{MHz}
\end{equation}

If the dark photon mass is in the neV mass range, corresponding to some MHz in frequency, then the interaction of a dark photon with the ionospheric plasma enhances production of real photons. However, when the dark photon mass is in the eV range or above, the conversion in the ionosphere is off resonance and therefore
smaller \cite{Dubovsky:2018arxiv}.

Radiation flux and power due to dark photon to photon conversion can vary widely depending on kinetic mixing, temporary flux enhancement due to gravitational focusing, resonance conditions, variability of local degree of ionization, and local dark matter density (see \cite{Gelmini:2020arxiv}).

For a $\approx$neV dark photon the coherence effects are determined by level crossings depending on the profile of ionization column as a function of altitude. Using a Chapman electron density profile given by ~\cite{Stankov:2003}

\begin{equation}
\label{eq:elecDensProf}
n_{e}(z )= n_{max} e^{\frac{1}{2}[1-\frac{z-h_{max}}{H}}-e^{(-\frac{z-h_{max}}{H})]}
\end{equation}

with $n_{max}=10^{12} \mbox{electrons} / \mbox{m}^3, H = 100~\mbox{km}, h_{max} = 300~\mbox{km}$, we calculate a scale factor of $|\frac{d\,log( \omega_{plasma}(z)^2)}{dz}|^{-1}\approx 10^4$ for resonant conversion in the the ionosphere~\cite{McDermott:2020arxiv,Garcia:2020}.

For a kinetic mixing value of $\epsilon=10^{-12}$, a dark matter density of $\rho=0.3 \, \mbox{GeV}/\mbox{cm}^{3}$, an antenna effective area of about $1 \, \mbox{m}^2$, and for a very modest gravitational amplification of only 100 from the Sun or solar system bodies, the power received in the MHz frequency range at stratospheric altitude is estimated as $\sim 5\times 10^{-15}$~W~\cite{Horns:2012arxiv}. 
The noise at these frequencies is mainly determined by system temperature and bandwidth. The system temperature includes sky noise and receiver noise. For a nominal system temperature of 5000~K, a bandwidth of 1~kHz, the noise power is $P_{noise}\approx 7\times 10^{-17}$~W yielding almost a factor of 100 signal to noise ratio. For weaker signals shielding techniques could be applied to reduce the receiver noise~\cite{Carr:2001}.

For an eV range dark photon the ionospheric plasma is very diluted and the conversions to photons are non-resonant, approximately like those in vacuum. Using power equation from \cite{Gelmini:2020arxiv} we estimate the number of photons from dark photons to be from about $10^{-6}-10^{4}\,\mbox{photons}\cdot\mbox{cm}^{-2}\cdot\mbox{s}^{-1}$. 

The sensitivity of an eV range measurement depends on the capabilities of detectors to measure the modulation in the presence of daytime and/or nighttime background. 
In the 150-200 nm range the dominant contribution of photon flux is from the Sun. Solar irradiance varies approximately linearly over the mentioned wavelength range from about $0.1\,\mbox{mW}\cdot\mbox{m}^{-2}\cdot\mbox{nm}^{-1}$ to $10 \,\mbox{mW}\cdot\mbox{m}^{-2}\cdot \mbox{nm}^{-1}$ at 
the top of the atmosphere, which then, at an altitude of $\approx 45$ km, is reduced by a factor $e$, corresponding to one optical depth ~\cite{Meier:1997AP}. 
Different absorption bands in this wavelength region and at these altitudes result in a spectral flux ranging from $10^{8}-10^{11}\,\mbox{photons}\cdot\mbox{cm}^{-2}\cdot\mbox{s}^{-1} \cdot\mbox{nm}^{-1}$~\cite{Meier:1997AP}. For daytime measurements, in order to suppress the solar light, a coronagraph could be used with a typical suppression of $10^6$ resulting in a photon flux of $5\times 10^{3} -5\times 10^{6} \,\mbox{photons}\cdot\mbox{cm}^{-2}\cdot\mbox{s}^{-1}$ over this range of wavelengths.  If daytime measurement are made, this is the main photon background.

Missions have made night-time measurements in low Earth orbit ($\approx 600 \, \mbox{km}$) in this energy range \cite{Henry:2015ApJ}.  They report a baseline spectral flux of $\approx 10^{4}\,\mbox{photons}\cdot\mbox{cm}^{-2}\cdot\mbox{s}^{-1}\cdot\mbox{nm}^{-1}$ which is primarily from diffuse galactic and extragalactic light and airglow of the atmosphere.  Reducing the flux by a factor of one optical depth to scale it at altitudes of 45 km, integrating over the solid angle, and considering different absorption bands in this wavelength region, the estimated cosmic noise level is $500-3\times 10^{5}\,\mbox{photons}\cdot\mbox{cm}^{-2}\cdot\mbox{s}^{-1}$ for the 150 nm -200 nm wavelength range. This would be the main background for night-time measurements.  

The overall signature would be a modulated 6-8\,eV photon flux or an RF power in the MHz range, concurrent with some planetary correlation or simultaneous temperature excursions in the stratosphere. Thus, at the beginning, the primary detector choices would be UV photon detectors and low noise radio receivers. Other detectors could be added on to probe other energy bands, wider fields of view, direction discrimination, and ancillary parasitic detectors as payload limit allows. 

Small payloads could be sent out first in weather balloon flights lasting hours to measure the day and night background and establish the best detector combinations for these measurements. This could then be followed by bigger and longer duration balloon missions with multi-purpose detectors.  

\begin{figure}[h]
\centering\includegraphics[width=1.0\linewidth]{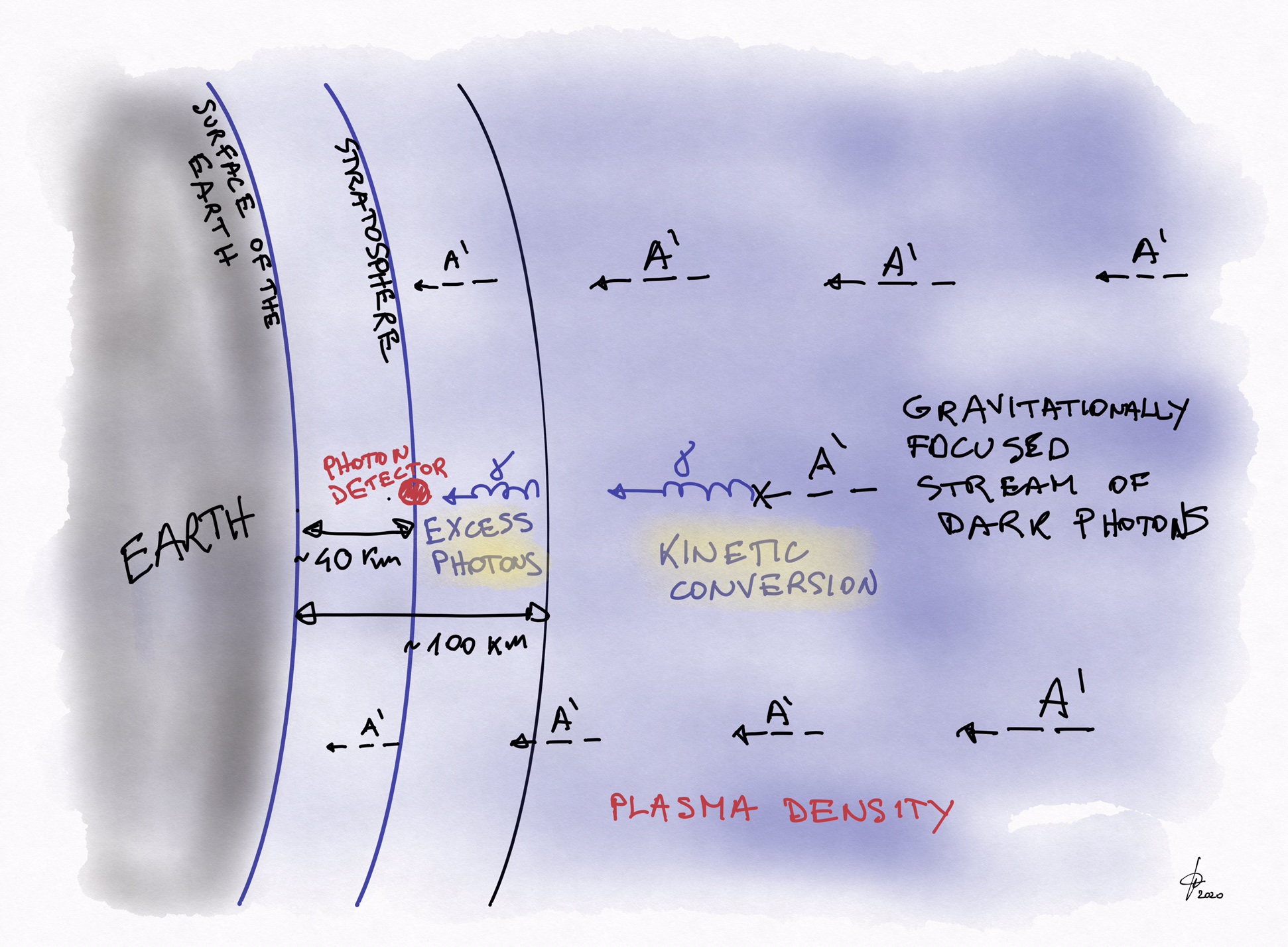}
\caption{Cartoon illustration of the direct search concept (not to scale). A gravitationally focused stream of dark photons partially converts into real photons in a region farther that $\approx$100\,km above the surface of the earth, where the plasma density has the resonant value. Converted photons are then absorbed in the upper stratosphere, $\approx$40\,km above the surface, causing the observed local temperature increase \cite{Zioutas:2020PDU}. A photon detector placed in the upper stratosphere could directly measure excess photons coming from converted dark photons (A').}
\end{figure}

\section{Possible detectors}
\label{S:3}

Balloon flights carrying a payload of several hundred kilograms in the stratosphere can last for days, providing the opportunity to attempt direct detection of DM as outlined above.
The ideal device fitting the proposed search would be a highly efficient, wide sensitive area, low noise photon detector covering a 4$\pi$ Field-Of-View (FOV), divided at least in two separate halves. The detector sections pointing away from Earth would search for variations in the incoming photon flux in the energy range of interest. The opposite sectors would act as background monitors.
For the detection of 6-8 eV photons from converted dark photons, a promising option would be using Silicon Photomultipliers (SiPMs). These detectors are solid-state devices,
Commercially available SiPMs can be arranged in arrays to cover active areas of several cm$^2$, and identical arrays could be combined with light guides to effectively cover large opposite portions of the sky. This would allow the pairwise comparison between detectors looking skywards and earthwards.
A preliminary test payload could also be equipped with UV sensitive photo-diodes for complementary monitoring.
In a possible payload configuration, SiPMs could be combined with RF receivers, such as those developed for radio astronomy, to provide sensitivity in the MHz range. For instance receivers with a dipole-like antenna developed for The Long Wavelength Array~\cite{Henning:2010arxiv} would suffice.
Balloon payloads could also be complemented with unconventional state-of-the-art detectors such as ultra-high sensitivity opto-mechanical force sensors \cite{Karuza:2016pdu}. The force sensors could be operated either in the dark mode where it would look for interactions with dark sector particles with its surface, or light mode where also the radiation pressure from photons will be sensed \cite{Arguedas:2019pdu}.

\section{Conclusions}
\label{S:4}

We have presented here a novel proposal to search for Dark Sector constituents converting into Standard Model particles, which then cause observable effects at upper stratospheric altitudes. The proposal relies on the idea of placing detectors in the stratosphere.

Although we focus on the example of dark photons kinetically converting into real photons, where the mixing can also be resonantly enhanced by the intervening ionospheric plasma, the search could be extended to other candidates from the Dark Sector by envisioning a suitable multi-purpose detector. The proposed search could also include instruments in orbit.

The other important feature is that the signature of Dark Sector conversion must show a dependence on the relative positions of the planets in our solar system. Conventional cosmic rays are mostly relativistic and therefore the solar system cannot exert a significant gravitational impact on their trajectories, certainly not on photons.
The new search strategy proposed in this work is fully complementary to underground experiments. Even surface detectors could not perform this type of search for some Dark Sector candidates as they may be shielded by the Earth’s atmosphere with about $10^3 \, \mbox{g} \cdot \mbox{cm}^{-2}$.

\section{Acknowledgements}
\label{S:5}
For M. Maroudas, this research is co-financed by Greece and the European Union (European Social Fund- ESF) through the Operational Programme ‘‘Human Resources Development, Education and Lifelong Learning’’ in the context of the project ‘‘Strengthening Human Resources Research Potential via Doctorate Research’’ (MIS-5000432), implemented by the State Scholarships Foundation (IKY). This work has been partially supported by University of Rijeka, Croatia, grant number 18-126.

\bibliographystyle{elsarticle-num-names}
\bibliography{DM_stratosphere.bib}

\end{document}